\documentclass{article}

\usepackage{arxiv}

\usepackage[utf8]{inputenc} % allow utf-8 input
\usepackage[T1]{fontenc}    % use 8-bit T1 fonts
\usepackage{hyperref}       % hyperlinks
\usepackage{url}            % simple URL typesetting
\usepackage{booktabs}       % professional-quality tables
\usepackage{amsfonts}       % blackboard math symbols
\usepackage{nicefrac}       % compact symbols for 1/2, etc.
\usepackage{microtype}      % microtypography
\usepackage{lipsum}
\usepackage{natbib}
\usepackage{amsmath}
\usepackage{listings}
\usepackage{wrapfig}
\usepackage{graphicx}
\graphicspath{ {images/} }
\usepackage{caption}
\usepackage{subcaption}
\usepackage{algpseudocode}
\usepackage{algorithm}
\title{\textsc{SeismiQB} ~--- a novel framework for deep learning with seismic data}

\author{
  Alexander Koryagin \\
  Data Analysis Center\\
  Gazpromneft\\
  Saint Petersburg \\
   \And
 Roman Khudorozhkov \\
  Data Analysis Center\\
  Gazpromneft\\
  Saint Petersburg \\
  \And
    Darima Mylzenova \\
  Data Analysis Center\\
  Gazpromneft\\
  Saint Petersburg \\
   \And
 Sergey Tsimfer\\
  Data Analysis Center\\
  Gazpromneft\\
  Saint Petersburg \\
}

\begin{document}
\maketitle

%%%%%%%%%%%%%%%%%%%%%%%%%%%%%%%%%%%%%%%%%%%%%%%%%%%%%%%%%%%%%%%%%%%%%%%%%%%%%%%%%%%%%%
\begin{abstract}
In recent years, Deep Neural Networks were successfully adopted in numerous domains to solve various image-related tasks, ranging from simple classification to fine borders annotation. Naturally, many researches proposed to use it to solve geological problems. Unfortunately, many of the seismic processing tools were developed years before the era of machine learning, including the most popular \textsc{SEG-Y} data format for storing seismic cubes. Its slow loading speed heavily hampers experimentation speed, which is essential for getting acceptable results. Worse yet, there is no widely-used format for storing surfaces inside the volume (for example, seismic horizons). 

To address these problems, we've developed an open-sourced \textsc{Python} framework with emphasis on working with neural networks, that provides convenient tools for (i) fast loading seismic cubes in multiple data formats and converting between them, (ii) generating crops of desired shape and augmenting them with various transformations, and (iii) pairing cube data with labeled horizons or other types of geobodies.
\end{abstract}

%%%%%%%%%%%%%%%%%%%%%%%%%%%%%%%%%%%%%%%%%%%%%%%%%%%%%%%%%%%%%%%%%%%%%%%%%%%%%%%%%%%%%%
\section{Introduction}

Seismic cubes contain crucial information about the structure of subterranean depth and multiple steps of seismic interpretation workflow rely on that type of data: segmenting facies, horizon detection and building velocity models to name a few. Training machine learning models requires a lot of iterations over the data, and speed of data loading is of utmost importance. At the same time, there is a lot of other variables to tinker with: for example, in some tasks, it is not uncommon to convert cube of amplitudes into cube of instantaneous phases or frequencies. All of this sets the necessity for convenient tools for fast prototyping.

To this end, we provide \textsc{SeismiQB} ~--- \textsc{Python} framework consisting of multiple components that allow for faster data loading as well as applying various transforms directly to values. It is also capable of working with 2D surfaces inside 3D volume: for example, seismic horizons, or with 3D bodies inside the cube: rivers estuaries can be seen as such case. Another feature of the framework is its ability to define sophisticated neural networks with just a few lines of code.

This paper sequentially describes various parts of the framework, both from a semantic and technical point of view; it presents key features of working with cubes and surfaces, as well as thoroughly walks through procedure of data generation.

%%%%%%%%%%%%%%%%%%%%%%%%%%%%%%%%%%%%%%%%%%%%%%%%%%%%%%%%%%%%%%%%%%%%%%%%%%%%%%%%%%%%%%
\section{Library features}

\subsection{Seismic cubes}
Individual cubes are stored in \textsc{SEG-Y} \cite{segy} format, which is, essentially, a container for individual traces. Trace is a 1D array of constant size with each value being an amplitude of reflected signal, provided with additional meta-information about location of the trace (both worldwide and in local coordinates), its internal and external parameters. Meta information allows to arrange traces on a rectangular grid of cross-lines and inlines, creating a 3D array of seismic data. Note that every cube is shot with completely  different equipment, and therefore have varying ranges of values, needing additional normalization. Meta-information also differs: some parameters present in one cube is absent in the others, and vice versa. Across all of the cubes, though, the following fields are consistent: world-wide coordinates, local coordinates, time-delay and sample rate. 

The size of the cubes is also different: it can vary in the range between few to hundreds of gigabytes, with different sizes of spatial dimensions and depth.

As we discuss in later sections, in order to train a neural network we must be able to cut data from the cube as fast as possible. Unfortunately, \textsc{SEG-Y} data format is not well-suited for this: it does not take advantage of knowing the shape and location of the crop to cut in advance. To overcome this limit, we convert all of our cubes into \textsc{HDF5} \cite{hdf5} file format, that allows slicing data along the desired dimension with great speed. To be able to make fast slices along every dimension, we physically store three copies of the same data: one that has inlines as the first axis, one that has cross-lines as the first axis, and the depth-oriented one. This essentially allows us to benefit from disk usage/loading speed trade-off: even the tripled demand on disk usage is negligible for modern systems. Meta information about traces is stored in additional fields.

Class \texttt{SeismicGeometry} provides key functionality of working with both \textsc{SEG-Y} and \textsc{HDF5} formats, as well as tools for converting from one to another. It also allows to check consistency of the cube so that meta-information and actual layout of the data do not contradict with each other. Note that this class only infers key information about cube: it does not actually load into memory more than one trace at a time.

\texttt{SeismicCubeset} allows to index multiple cubes at once. Due to completely different ranges of amplitude values in cubes, we implemented a tool to scale them into $[0, 1]$ range: otherwise, the inter-cube generalization would be too hard. It is also well-known that applying other transforms to the values can be of great help for some of the seismic-related tasks: for example, converting seismic volume into a cube of instantaneous phases can help to track seismic reflections. This can be used to incorporate prior knowledge into learned models and can be easily done in our library.

\subsection{Hand-labeled horizons}\label{GT}
In most tasks, we work in a supervised way, and each seismic cube is paired with multiple horizons or other entities like river estuaries. Each of them is either 2D surface or 3D volume inside a 3D cube of values and stored as a cloud of $(x, y, t)$ points with axis corresponding to cross-lines, inlines and time. Throughout the article we interchangeably use words \textit{time}, \textit{depth} and \textit{height}, and the bigger values correspond to a deeper location of the point.

As the labeled entities take a very small fraction of cube volume, we need to be able to cut crops near them during model training. \texttt{SeismicCubeset} can create a tool called \textit{sampler} for sampling such points according to the distribution of the labeled points inside the cube, and one can easily modify it to change its behavior: for example, to generate points only from the half of the cube (along any of the dimensions), or to use only every i-th slide of the data (again, along any of the dimensions). This is essential to have a fair evaluation of predictive models: we use this capability to split data into train/validation sets, and it can be used to create different learning settings.

\begin{wrapfigure}{r}{0.3\textwidth}
  \vspace{-10pt}
  \centering
  \includegraphics[width=.3\textwidth]{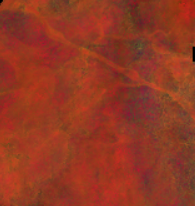}
  \captionof{figure}{View from above can help identifying river estuaries}
  \vspace{-40pt}
  \label{fig:river}
\end{wrapfigure}

As the framework is capable of working not only with the hand-made labels, but with auto-detected ones, we also provide methods to filter, smooth out, and apply other functions to point-cloud. There are also a lot of functions to visualize particular horizon from different points of view: for example, looking at cube values along the labeled surface can be used to track various 3D bodies, as shown at Image \ref{fig:river}.

\subsection{Data-feeding procedure}
Despite immense progress in modern processing units (CPU, GPU, TPU), the amount of information even in a single cube still poses a significant challenge: it hardly fits into memory constraints, so we need to sequentially generate crops of data to train the model on. As was noted before, we use the shape of the crop to cut to infer the fastest axis to use as the first one: that can result in up to 30 times faster loading of badly oriented volumes. To leverage the capabilities of present-day systems with multiple CPU's, we load crops in multiple threads, while training is usually performed on GPUs. 

Naturally, our framework also allows to apply various transforms and augmentations to both data and labels, thus enhancing model performance and making it robust to distortions; an example of this can be seen at [IMG REF]. 

Most of this workflow is encapsulated into \texttt{SeismicCropBatch}: instances of this class use \texttt{SeismicCubeset}'s \textit{sampler} to generate points to cut data from, then load actual values from cubes, add hand-made labels specific to the task, and, optionally, apply additional augmentations.

Our library heavily relies on \textsc{BatchFlow} in order to define even the most sophisticated neural networks with just a few lines of code. There are also implementations for various popular architectures, ranging from simple \textsc{ResNet} \cite{resnet} and \textsc{UNet}-based \cite{unet} models to state-of-the-art \textsc{EfficientNet} \cite{efficientnet} and \textsc{DeepLab} \cite{deeplab}. This library also allows us to train models on multiple GPU's and provides a tool for fine memory control during training, which is essential when working with such enormous amounts of data.

%%%%%%%%%%%%%%%%%%%%%%%%%%%%%%%%%%%%%%%%%%%%%%%%%%%%%%%%%%%%%%%%%%%%%%%%%%%%%%%%%%%%%%
\section{Conclusion}
In this paper, we present a novel framework for working with seismic data. It allows dynamically cut crops of varying shapes from seismic volumes in various data formats, apply transforms to the values and add labels to train on. We plan to build more results based on both our library and dataset by creating individual predictive models for tackling seismic tasks like horizon detection, facies segmentation and so on.

\bibliographystyle{unsrt}
\bibliography{references}

\end{document}